\documentclass[%
preprint,
superscriptaddress,
showpacs,
preprintnumbers,
graphicx,
floatfix,
byrevtex,
amsfonts,
amsmath,
amssymb,
bibnotes,
aps,
]{revtex4-1}

\usepackage{float,graphicx}
\usepackage{dcolumn}
\usepackage{bm}
\usepackage[mathlines]{lineno}
\usepackage{ulem}
\usepackage{color}
\usepackage[version=3]{mhchem}
\usepackage{mathrsfs,braket}

\newcommand*\hmote{2H$-\mathrm{MoTe_2}$\ }
\newcommand*\tmote{1T'$-\mathrm{MoTe_2}$\ }

\begin{document}
\preprint{} 

\title{Photo-induced Tellurium segregation in $\mathbf{MoTe_2}$}

\author{Takumi Fukuda}
\email{s2130057@s.tsukuba.ac.jp}
\affiliation{Department of Applied Physics, Graduate school of Pure and Applied Sciences, University of Tsukuba, 1-1-1 Tennodai, Tsukuba 305-8573, Japan}
\author{Ryota Kaburauchi}
\affiliation{Department of Applied Physics, Graduate school of Pure and Applied Sciences, University of Tsukuba, 1-1-1 Tennodai, Tsukuba 305-8573, Japan}
\author{Yuta Saito}
\affiliation{Device Technology Research Institute, National Institute of Advanced Industrial Science and Technology (AIST), Tsukuba Central 2, 1-1-1 Umezono, Tsukuba 305-8568, Japan}
\author{Kotaro Makino}
\affiliation{Device Technology Research Institute, National Institute of Advanced Industrial Science and Technology (AIST), Tsukuba Central 2, 1-1-1 Umezono, Tsukuba 305-8568, Japan}
\author{Paul Fons}
\affiliation{Department of Electronics and Electrical Engineering, School of Integrated Design Engineering, Keio University, 4-1-1 Hiyoshi, Kohoku district, Yokohama city 223-8521, Japan}
\author{Keiji Ueno}
\affiliation{Graduate School of Science and Engineering, Saitama University, Saitama 338-8570, Japan}
\author{Muneaki Hase}
\email{mhase@bk.tsukuba.ac.jp}
\affiliation{Department of Applied Physics, Graduate school of Pure and Applied Sciences, University of Tsukuba, 1-1-1 Tennodai, Tsukuba 305-8573, Japan}

\date{\today}

\begin{abstract}
We report on the coherent phonon spectroscopy of polymorphic $\mathrm{MoTe_2}$ single crystals using a femtosecond-pulsed laser to investigate the relationship between structural phase transitions and photo-thermal effects induced by high-density laser excitation. 
Even when a femtosecond pulsed laser was used, which generally induces fewer heat accumulation effects than the case for a CW laser, tellurium segregation was observed in coherent phonon spectra with increasing excitation density, while no structural phase transition among polymorphs of $\mathrm{MoTe_2}$ was observed. 
The use of higher photon-energy excitation, however, was found to suppress tellurium segregation.
\end{abstract}

\maketitle


\section{Introduction}
\begin{figure}
	\centering
	\includegraphics[width=16cm]{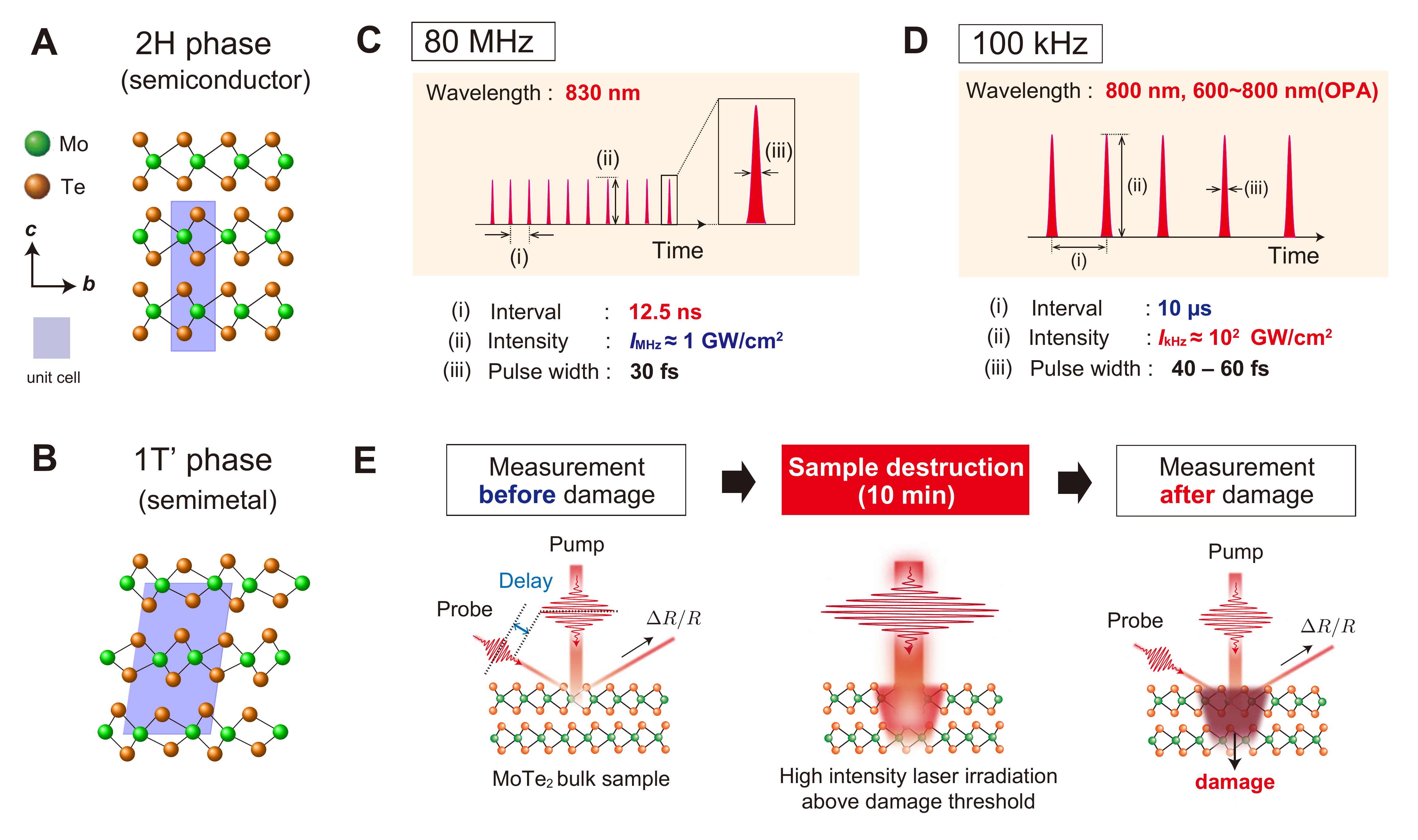}
	\caption{
	Schematic crystal structures of \textbf{A}. $\mathrm{2H-MoTe_2}$ and \textbf{B}. $\mathrm{1T'-MoTe_2}$.	\textbf{C} and \textbf{D}. Comparison of the experimental conditions between 80 MHz and 100 kHz repetition pulsed lasers. \textbf{E}. Experimental schemes of the present reflective-type pump-probe spectroscopy before and after optical sample damage.
	}
	\label{intro}
\end{figure}

\ Structural polymorphism of transition-metal dichalcogenides (TMDCs), a class of two-dimensional (2D) layered materials composed of a transition-metal and two chalcogens, has been attractive from a fundamental physics viewpoint for the study of non-equilibrium structural
dynamics as well as for memory applications as a phase-change material \cite{mannebach2017dynamic, zong2018ultrafast,krishnamoorthy2019optical, sie2019ultrafast, zhang2019light, fukuda2020ultrafast}. 
$\mathrm{MoTe_2}$, a well-known TMDC, is a representative polytype and exhibits three distinct crystal structures: a semiconducting $\mathrm{2H}$ (FIG. \ref{intro}A), a semimetallic $\mathrm{1T'}$ (FIG. \ref{intro}B), and a Type-II Weyl semimetallic $\mathrm{T_d}$ phase \cite{sun2015prediction, deng2016experimental}.
The $\mathrm{2H}$ phase is stable at room temperature (300 K), the $\mathrm{1T'}$ phase is metastable at room temperature and the $\mathrm{T_d}$ phase is a low temperature ($< 250\ \mathrm{K}$) polymorph of the $\mathrm{1T'}$ phase. 
Reversible switching between the semiconducting and metallic phases of $\mathrm{MoTe_2}$ by illumination with light is of particular interest for the development of novel device fabrication methods, such as an atomic scale phase-patterning technology with the low-dimensional structures for the fabrication of atomically thin high-performance integrated circuits \cite{cho2015phase,tan2018controllable,kang2021phase}.
Illumination with light also offers alternative technique to induce phase transitions over other methods such as temperature \cite{Keum2015bandgapopening}, electronic doping \cite{wang2012electronics}, strain \cite{song2015room}.
Theoretical studies suggest that high-density electronic excitation of the semiconducting $\mathrm{2H}$ phase by light pulses with a photon energy above the band gap can induce a structural phase transition into a metallic $\mathrm{2H^*}$ or semi-metallic $\mathrm{1T'}$ phase \cite{kolobov2016electronic, krishnamoorthy2018semiconductor,peng2020sub}.
However, experimental studies of light-induced structural phase transitions remain controversial: when a strong continuous-wave (CW) laser is used, tellurium segregates from $\mathrm{MoTe_2}$ via photo-thermal or heat-accumulation effects \cite{sakanashi2020investigation}, as surface oxidation in the presence of  elevated temperature even under vacuum conditions \cite{ueno2015changes}. 
On the other hand, there have been few studies of light-induced structural phase transitions in $\mathrm{MoTe_2}$ using femtosecond pulses, for which photo-thermal effects are expected to be significantly less than the case for CW laser irradiation since femtosecond electronic excitation occurs before thermal effects appear \cite{sundaram2002inducing}. 
Here, we explore the lattice dynamics of $\mathrm{2H}$- and \tmote samples monitored by coherent phonon spectroscopy under high-intensity irradiation with a 30-45 fs pulsed laser at ambient conditions (300 K) to gain insights into the possibility of controlling the structural polymorphism of $\mathrm{MoTe_2}$ by light in the presence of minimal thermal effects. \par

\section{Experimental}
\ The flake form $\mathrm{MoTe_2}$ bulk single crystals used were grown by the flux zone method for the 2H phase (commercially available from \textit{2D Semiconductors}, USA), and chemical vapor transport (CVT) for the 1T' phase \cite{ueno2015introduction}.
The thickness of the samples used was about $100\ \mathrm{\mu m}$.
Based on density functional theory (DFT) simulations, the optical penetration depth was estimated to be $\approx \mathrm{45\ nm}$ for the photon energy of $\mathrm{1.5\ eV}$ ($\lambda \approx $ $\mathrm{800\ nm}$) and $\approx \mathrm{25\ nm}$ for $\mathrm{2.0\ eV}$ ($\lambda \approx $ $\mathrm{600\ nm}$) for both the 2H and 1T' phases, which suggests there is negligible photon-energy dependence. 
Since the sample thickness is much thicker than the optical penetration depth, the effects of the substrate or the sample thickness dependence is negligibly small in our study.
Coherent phonons, whose generation process is traditionally described as resonant or non-resonant impulsive stimulated Raman scattering (ISRS) \cite{stevens2002coherent} or displacive exciation of coherent phonon (DECP) \cite{zeiger1992theory}, were detected by a degenerate reflective-type optical pump-probe technique that uses the same wavelength for the pump and probe pulses. 
We used two types of femtosecond light sources to evaluate the resulting photo-thermal and heat-accumulation effects as shown in FIG. \ref{intro}C and D : one is 30-fs pulse duration, 830 nm ($= 1.49\ \mathrm{eV}$) central wavelength with an 80-MHz repetition rate from a Ti:Sapphire oscillator, with the other had a duration of 45 fs, 100 kHz repetition rate, a 800-nm ($= 1.55$-$\mathrm{eV}$) central wavelength  generated with a regenerative amplifier and an optical parametric amplifier (OPA) to allow variation of the central wavelength of the light. The OPA generated light between $1200\ \mathrm{nm} \sim 1600\ \mathrm{nm}$. Both $600\ \mathrm{nm} \ (= 1.78\ \mathrm{eV})$ and $700\ \mathrm{nm} \ (= 2.06\ \mathrm{eV})$ light is available via second harmonic generation (SHG) using a BBO crystal under phase-matching conditions. 
The high repetition 80-MHz laser delivers femtosecond pulses with an interval of 12.5 ns and a peak intensity of $I_{\mathrm{MHz}} \approx 1\ \mathrm{GW/cm^2}$.  
On the other hand, the low repetitive 100-kHz laser delivers femtosecond pulses with an interval of $10\ \mathrm{\mu s}$ and a peak intensity of $I_{\mathrm{kHz}} \approx 10^2\ \mathrm{GW/cm^2}$.  
It is noted that the pump and probe light is refocused to a spot with a diameter estimated to be about $\mathrm{20\ \mathrm{\mu m}}$ on the sample surface.
Taking into account the fact that the photo-induced temperature rise accumulates on a  sub-microsecond time scale, photo-thermal effects induced by the high repetition pulsed laser are larger than those induced by the low repetition light source, while the excitation density per light pulse of the low repetitive one is larger than the high repetitive one.
As is schematically described in FIG. \ref{intro}E,  to investigate for a possible structural phase transition, we performed measurements of coherent phonon signals and took images of sample surfaces by optical microscopy before and after 10-minutes laser irradiation above the damage threshold without changing the irradiation spot location. 
It is noted that the measurements were conducted under an ambient condition and the pump fluence used for before-and-after measurements was set to an intensity lower than the damage threshold.
By comparing the phonon spectra and the optical contrasts, photo-induced structural changes can be evaluated. The figures used in this article were generated using Matplotlib \cite{Hunter:2007}. \par

\begin{figure}
	\centering
	\includegraphics[width=8cm]{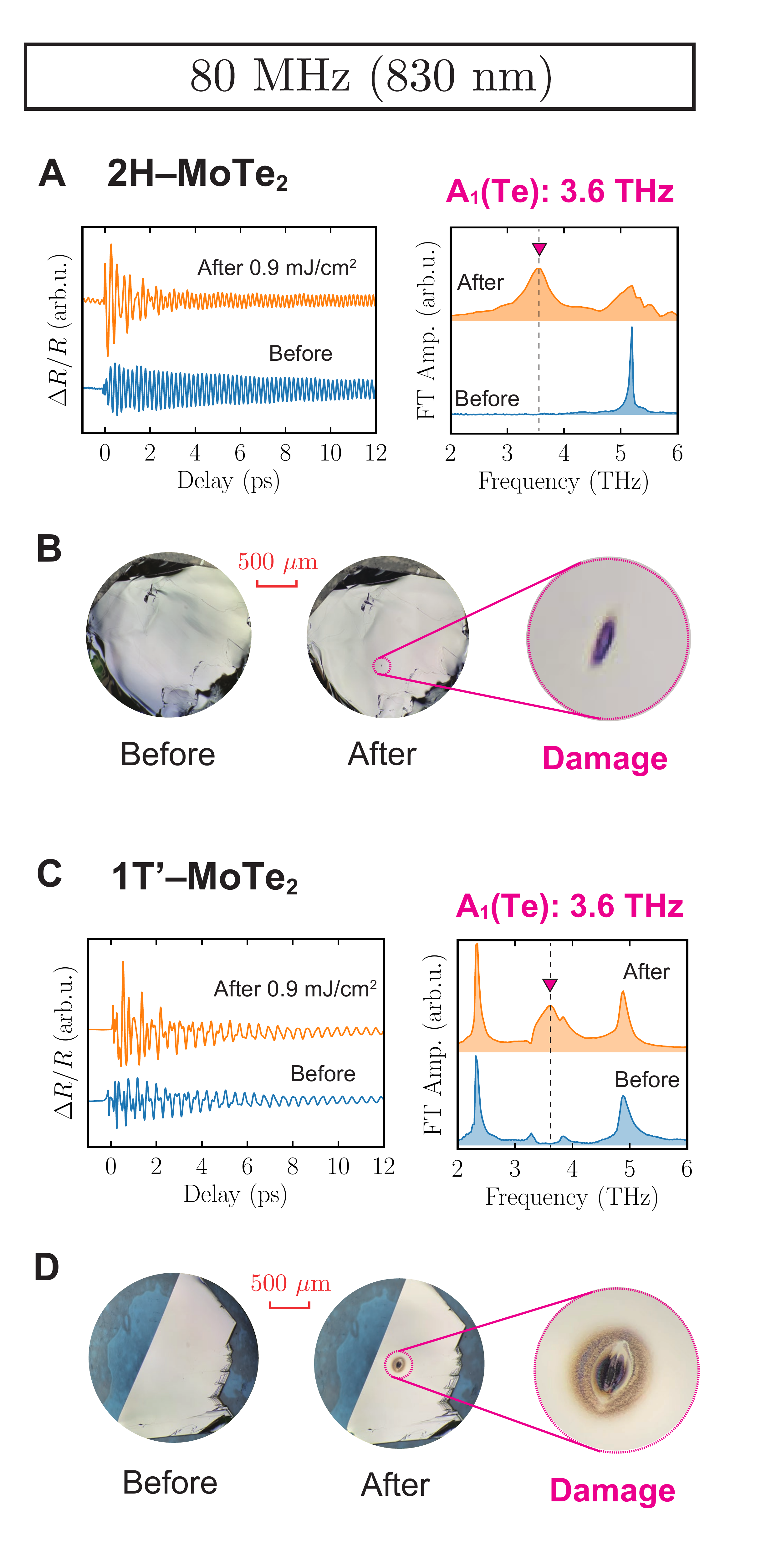}
	\caption{\textbf{80-MHz experiments.} 
	Experimental coherent phonon signals, phonon spectra (\textbf{A}:\ \hmote, \textbf{C}:\ \tmote) and optical microscopy images of the sample surfaces (\textbf{B}:\ \hmote, \textbf{D}:\ \tmote) before and after laser irradiation ($0.9\ \mathrm{mJ/cm^2}$) near the destructive threshold using a 80-MHz with 830-nm pulsed laser.
	The dashed lines in the FT spectra in \textbf{A} and \textbf{C} indicate 3.6 THz, which is the $A_1$ mode frequency of tellurium.
	}
	\label{80MHz_results}
\end{figure}

\begin{figure}
	\centering
	\includegraphics[width=8cm]{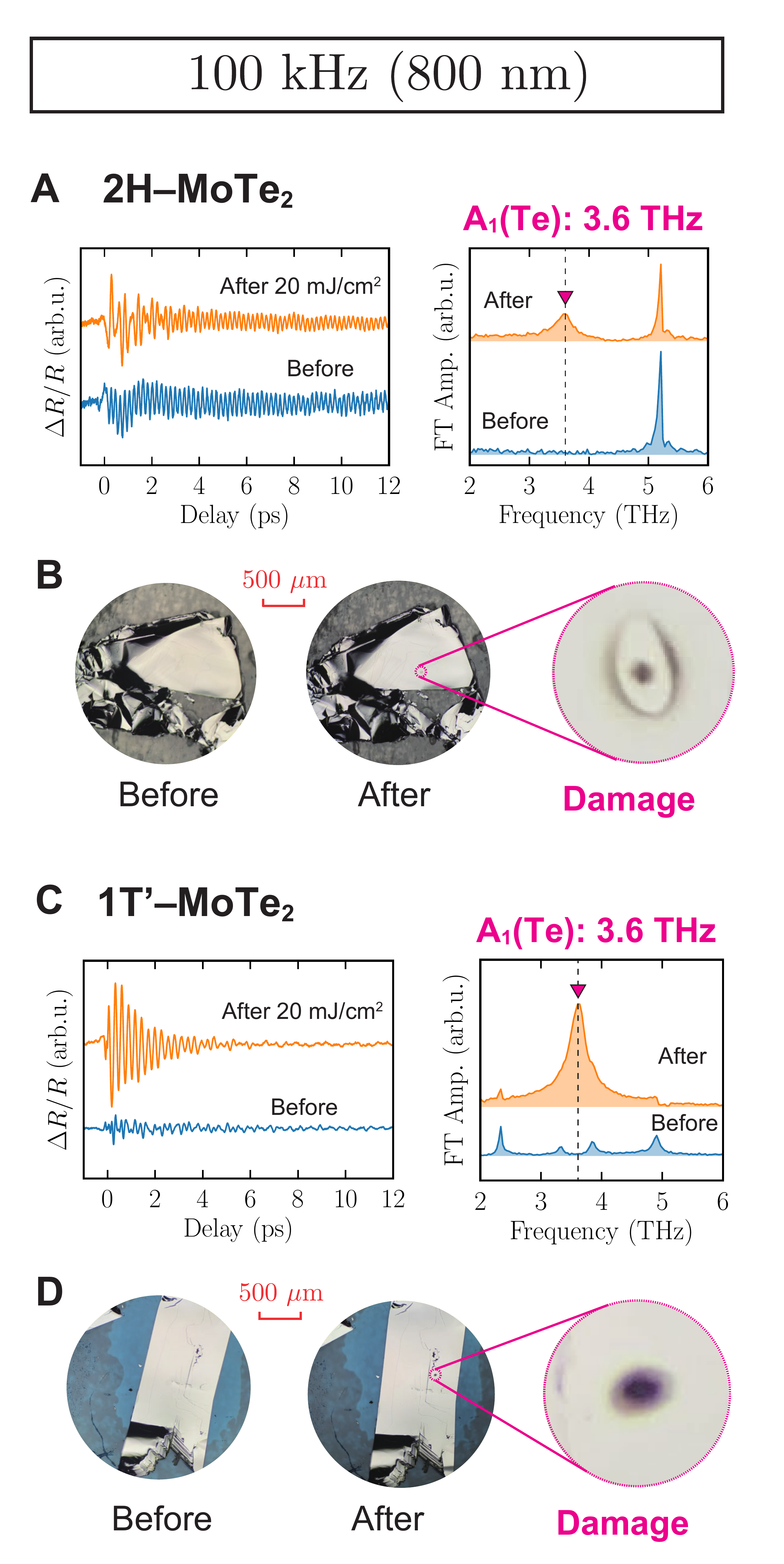}
	\caption{\textbf{100-kHz experiments.}  Experimental coherent phonon signals, FT phonon spectra (\textbf{A}:\ \hmote, \textbf{C}:\ \tmote) and optical microscopy images of the sample surfaces (\textbf{B}:\ \hmote, \textbf{D}:\ \tmote) before and after laser irradiation ($20\ \mathrm{mJ/cm^2}$) near the destructive threshold using a 100-kHz with 800-nm pulsed laser. 
	The dashed lines in the FT spectra in \textbf{A} and \textbf{C} indicate 3.6 THz, which is the $A_1$ mode frequency of tellurium.
	}
	\label{100kHz_results}
\end{figure}

\section{Results and discussion}

\ First, we show the coherent phonon signals before and after surface damage due to the use of the 80-MHz pulsed laser with $\lambda = 830\ \mathrm{nm}$.
To analyze the oscillatory components of the time-domain signals, we subtracted a background signal derived from electronic response using biexponential function.
The damage-threshold pump fluence for both the $\mathrm{2H-}$ and $\mathrm{1T'-MoTe_2}$ samples was found to be about $F_{\mathrm{th}(\mathrm{MHz})} = \mathrm{0.9\ mJ/cm^2}$ for the 80-MHz measurements, so that the pump fluence used was set to $0.3\ \mathrm{mJ/cm^2}$, a value much lower than $F_{\mathrm{th}(\mathrm{MHz})}$.
FIG.  \ref{80MHz_results}A, C show the coherent phonon signals and Fourier transformed (FT) phonon spectra observed in $\mathrm{2H-}$ and $\mathrm{1T'-MoTe_2}$ before and after sample damage.
Before laser irradiation, coherent phonon oscillations from both samples are visible for time ranges longer than 25 ps, indicating that the scattering terms such as electron-phonon and phonon-phonon interaction are quite small as expected based on the characteristic low dimensional structure of TMDCs.
From the FT spectra of the coherent phonon signals, we found one $A_{1g}$ mode ($5.1\ \mathrm{THz}$) for $\mathrm{2H-MoTe_2}$ and four $A_{g}$ modes ($2.3\ \mathrm{THz},\ 3.3\ \mathrm{THz},\ 3.9\ \mathrm{THz}$, and $4.9\ \mathrm{THz}$) for $\mathrm{1T'-MoTe_2}$, which are in good agreement with the previous reports of Raman and coherent phonon measurements on the 2H and 1T' phases \cite{sugai1982high, froehlicher2015unified, chen2016activation, zhang2016raman, ma2016raman, zhang2019light, fukuda2020ultrafast}.
It is noted that, since we carried out the experiment using an isotropic reflection geometry for the coherent phonon detection, anisotropic lattice vibration modes, such as $E$ modes, are expected to be difficult to detect in our measurements. 
After laser irradiation, we observed that a new vibrational mode at 3.6 THz\ ($\approx 120\ \mathrm{cm^{-1}}$) appeared for both samples, accompanied by ablation-like permanent damage patterns in the form of a black discoloring in optical contrast as can be seen in FIG. \ref{80MHz_results}B and D.
The emergent mode at 3.6 THz is considered to be the $A_{1}$ mode of tellurium (Te) \cite{hunsche1995impulsive, hunsche1996details, roeser2004optical, misochko2007coherent, kamaraju2010large} due to segregation of atomic Te from the $\mathrm{MoTe_2}$ samples as a result of surface melting induced by heat-accumulation effects of high repetitive pulsed laser irradiation. 
The Te segregation observed to date in coherent phonon spectra appearing at 3.6 THz has been occasionally observed in other telluride compounds, such as $\mathrm{Ge_2Sb_2Te_5}$ \cite{miller2016ultrafast}, $\mathrm{Sb_2Te_3}$ \cite{li2010coherent, norimatsu2015dynamics}, $\mathrm{CdTe}$ \cite{ishioka2006amplitude}, ZnTe \cite{shimada2014indication} as well as static-measurement studies on $\mathrm{MoTe_2}$  \cite{sakanashi2020investigation}.
On the other hand, additional phonon modes related to the different structural phases were not observed after the laser irradiation: the 2H–to–1T' and 1T'–to–2H phase transitions were not observed.
Notably, the 2H phase has an intrinsic vibrational mode at 3.6 THz manifested as the $E_{1g}$ mode \cite{sugai1982high,  froehlicher2015unified, caramazza2018first} and Te also has an $E^1$ mode at 4.2 THz \cite{Pine1971phononTellurium,du2017one}. However, as mentioned above, our detection scheme is, in principle, insensitive to anisotropic lattice vibrations so that the scenario of Te segregation is the most plausible mechanism to describe our results. 
For these reasons, no structural phase transitions and Te segregation with permanent sample damage was observed for both 2H– and $\mathrm{1T'–MoTe_2}$ samples at $< 1\ \mathrm{mJ/cm^2}$ irradiation with the use of the high repetition 80-MHz pulsed laser.

\ Second, we examined the effects of the low repetition 100-kHz pulsed laser with $\lambda = 800\ \mathrm{nm}$.
The damage-threshold fluence $F_{\mathrm{th}(\mathrm{kHz})}$ for both $\mathrm{2H-}$ and $\mathrm{1T'-MoTe_2}$ samples was found to be between $10$ to $\mathrm{20\ mJ/cm^2}$ for the 100-kHz measurements, which is surprisingly about ten times larger than the threshold for the high repetition 80-MHz measurements, $F_{\mathrm{th} (\mathrm{MHz})} < F_{\mathrm{th} (\mathrm{kHz})}$.
This observation can be considered to be a result of the reduced heat accumulation effects due to the reduced repetition rate. 
The pump fluence used for the coherent phonon measurements was $1\ \mathrm{mJ/cm^2}$, a value much lower than the threshold $F_{\mathrm{th}(\mathrm{kHz})}$, while the pump fluence leading to sample damage was $20\ \mathrm{mJ/cm^2}$.
Before laser irradiation, the coherent phonon oscillations and the associated FT phonon spectra for both samples were similar to those of the 80-MHz experiments, as can be seen in FIG. \ref{100kHz_results}A and C. 
After laser irradiation, although the spectral details are slightly different from those for the case of the 80-MHz experiments, the $A_{1g}$ mode of Te segregation (3.6 THz) appears along with permanent damage patterns and no additional phonon modes related to the structural phases in both samples were observed, as shown in FIG. \ref{100kHz_results}B and D.
These results suggest that, even under the use of low- and high-excitation pulsed lasers, Te segregation is inevitable, and light-induced structural phase transitions between 2H and 1T' phases are hard to realize under the current experimental conditions. \par

\ As discussed above, Te segregation is considered to be strongly promoted by photo-thermal effects due to heat accumulation as can be seen in the observed differences in the effects with the pulsed laser repetition rate.
Since the samples we used in the measurement were exposed to air, the sample is thought to be oxidized as a consequence of laser irradiation, leading to the formation of other compounds. 
A recent investigation has revealed that molybdenum oxides such as $\mathrm{MoO}_{3-x}$ are formed on the surface of $\mathrm{MoTe_2}$ due to elevated temperature \cite{ueno2015changes}.
We also confirmed oxide formation in the ablation-like damage region using scanning electron microscopy and energy dispersive X-ray spectroscopy (SEM-EDX) measurements.
As in the case here, the remaining Te atoms segregate to form Te-rich regions in the presence of surface oxidation leading to the formation of  $\mathrm{MoO}_{3-x}$ by laser irradiation.
To avoid the surface oxidation and Te segregation, oxidation prevention countermeasures such as surface protection or high vacuum conditions are required to further investigate and control potential light-induced phase transitions among the polymorphs of $\mathrm{MoTe_2}$.
From another perspective, the use of different wavelengths may also play an important role in light-induced phase transitions because higher photon energies may exceed the energy barrier of the free energy landscape \cite{kolobov2016electronic}. According to a theoretical study, the use of the critical photon energy of 633 nm (= 1.96 eV) with dense electronic excitation can lower the potential barrier between the monolayer 2H and 1T' phases, which is thought to be also applicable to the bulk crystals. \par

\begin{figure}
	\centering
	\includegraphics[width=8cm]{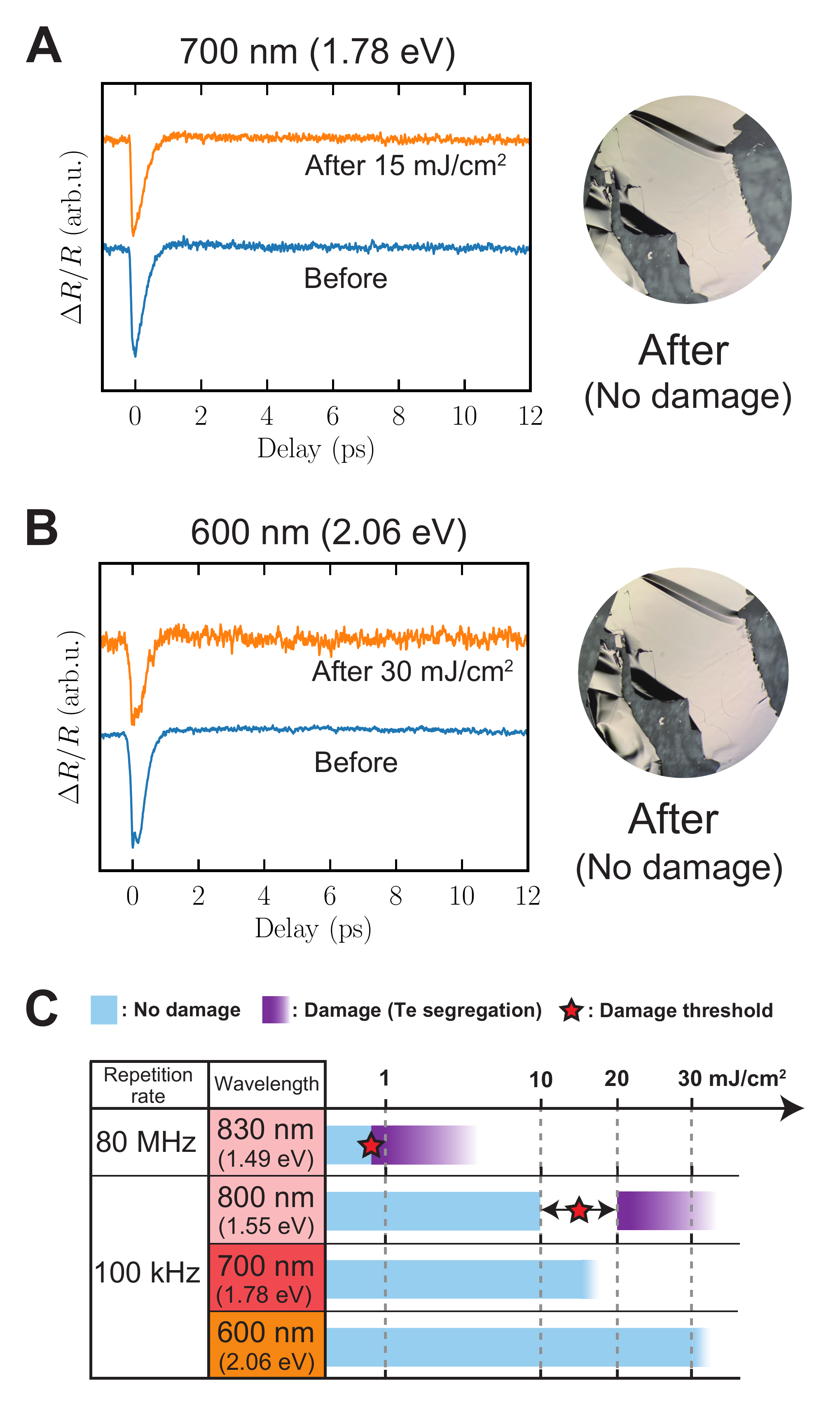}
	\caption{\textbf{Pump-probe measurements on} $\mathbf{2H-MoTe_2}$ \textbf{with an OPA and the relationship of the damage thresholds among the various pulsed-laser conditions.}
	$\Delta R/R$ signals and surface optical microscopy images after laser irradiation showing no damage patterns on  \textbf{A}. 700-nm (1.78-eV) light and \textbf{B}. 600-nm (2.06-eV) light.
	\textbf{C}. A schematic table summarizing the  damage-threshold study on repetition rates and laser wavelength.
	}
	\label{OPA_results}
\end{figure}

\begin{table}[b]
    \centering
    \caption{Absorption coefficients of the 2H phase ($\alpha_{\mathrm{2H}}$) and the 1T' phase ($\alpha_{\mathrm{1T'}}$) for 800 nm (1.55 eV), 700 nm (1.78 eV), and 600 nm(2.06 eV) calculated using DFT simulations.}
    \label{abcof}
    \begin{tabular}{c|ccc}
         & 800 nm (1.55 eV)& 700 nm (1.78 eV)& 600 nm (2.06 eV) \\ \hline \hline
        $\alpha_{\mathrm{2H}}$ & $2.3 \times 10^7\ (\mathrm{m^{-1}})$& $2.5 \times 10^7\ (\mathrm{m^{-1}})$ & $4.2 \times 10^7\ (\mathrm{m^{-1}})$ \\
        $\alpha_{\mathrm{1T'}}$ & $2.3 \times 10^7\ (\mathrm{m^{-1}})$ & $2.8 \times 10^7\ (\mathrm{m^{-1}})$& $3.6 \times 10^7\ (\mathrm{m^{-1}})$\\ \hline
    \end{tabular}
\end{table}

\ To test for the effects of the photon energy, we performed reflectivity change measurements using light with central maximums of 700 nm (= 1.78 eV) and 600 nm (= 2.06 eV) generated using an OPA.
FIG. \ref{OPA_results} A and B show the $\Delta R/R$ signals of $\mathrm{2H-MoTe_2}$ before and after laser irradiation with a fluence of $15\ \mathrm{mJ/cm^2}$ for 700 nm, and $30\ \mathrm{mJ/cm^2}$ for 600-nm, which are the maximum output for each wavelength for the OPA.
Even with the use of such high fluences, no surface damage was seen in the surface optical images for both the $\mathrm{2H}$ and $\mathrm{1T'}$ phases.
In this experiment, coherent phonons in the samples were invisible in $\Delta R/R$ signals except for the initial electronic response due to the pulse broadening of the OPA light compared with the 800-nm measurements or the wavelength dependence of coherent-phonon responses \cite{sato2014resonance}.
Although, due to the absence of coherent phonons, we cannot evaluate the structural dynamics under photon energies greater than 1.96 eV ($\lambda = 633\ \mathrm{nm}$), which is predicted to be a critical value for monolayer 2H-to-1T' phase transition \cite{peng2020sub}, the time-domain $\Delta R/R$ signals from the 600-nm measurement show little change before and after $30\ \mathrm{mJ/cm^2}$ irradiation, a value about twice as large as the damage threshold for 800-nm irradiation. 
That is, in addition to the repetition rate, the wavelength (photon energy) of the light source is possible related to the presence of the surface damage or the absence of Te segregation on the $\mathrm{MoTe_2}$ samples.
In general, the photon-energy dependence of ablation can be explained by absorption coefficient since changes in optical absorption directly contribute to photo-thermal or heat-accumulation effects.
The absorption coefficients $\alpha$ for 1.55 eV, 1.78 eV and 2.06 eV photon energies calculated using DFT simulations are listed in TABLE \ref{abcof}.
In both phases, we found the value of the absorption coefficient of 2.06 eV was lager than that of 1.55 eV, suggesting photo-thermal effects should be more significant at 2.06 eV than 1.55 eV. However, it is inconsistent with our observation that the Te segregation is suppressed for the photon energy of 2.06 eV, which is larger than 1.55 eV. This indicates that the photo-thermal effect induced by optical absorption cannot explain the observed supression of Te segregation for light with energies under 2.06 eV. 
Therefore, a systematic study of the structural dynamics of $\mathrm{MoTe_2}$ systems dependent on the photon energy with much higher intensities and much shorter pulse width is required in the future.
FIG. \ref{OPA_results}C shows a schematic table summarizing the relationship of the damage thresholds and Te segregation for $\mathrm{MoTe_2}$ to the repetition rates and laser wavelength.

\section*{Conclusion}%
\ We have investigated the structural changes induced by high-intensity femtosecond pulsed light above the sample-damage threshold. Reflective pump-probe coherent-phonon spectroscopy was carried out using two types of femtosecond pulsed lasers, a high repetitive 80-MHz with $\lambda = 830\ \mathrm{nm}$ and a low repetitive 100-kHz light source with a wavelength range of $\lambda = 800$, 700 and 600 nm, 
revealed that Te segregated on the surface of both phases leading to the observation of the $A_{1}\mathrm{(Te)}$ mode at 3.6 THz ($\approx \mathrm{120\ \mathrm{cm^{-1}}}$). In addition, ablation-like patterns were observed in the optical contrast, which are considered to be a consequence of the heat-accumulation effects that depend on the repetition rate.
When using higher photon-energy excitation, however, the sample damage was found to be significantly suppressed, indicating that Te segregation is also likely related to the laser wavelength.
Our results indicate that, even with the use of femtosecond pulsed light sources, Te segregation makes it difficult to realize the light-induced structural phase transitions between the 2H and 1T' phases. 
We suggest that the alternative sample structures to eliminate surface oxidation such as oxidation prevention layers, and the use of high vacuum or selective photon energies may be efficient ways to suppress Te segregation and realize the light-induced structural phase transitions.

\section*{Acknowledgement}%
The authors are grateful for financial supports from the Murata Science Foundation, JSPS KAKENHI (Grant Numbers. 17H02908, 18H01822 and 19H02619), CREST, JST (Grant Number. JPMJCR1875) and JSPS Research Fellowships for Young Scientists, Japan. 
We gratefully acknowledge Prof. T. Sekiguchi (R\&D Center for Innovative Material Characterization) for supporting SEM-EDX measurements. 

\bibliography{MoTe2_TeSegr}

\newpage

\end{document}